# A New Approach to Understanding Ontogenesis and The Theory of Aging

Lev Salnikov

Boston University Medical Center,

Boston, MA 02118

Send correspondence to Lev Salnikov, Email: leosalnikov@gmail.com

## ABSTRACT

This paper proposes an original theory of aging of multicellular organisms. The cells of multicellular organisms, in contrast to unicellular organisms, are burdened with a two- part genome: housekeeping and specialized (multicellular), responsible for ontogenesis and terminal differentiation. The two parts of the genome compete for limited adaptive resources thereby interfering with the ability of the house-keeping part of the genome to adequately perform reparative and adaptive functions in post mitotic cells. The necessity to complete the ontgenesis program, leads to increased activity of the multicellular components of the genome. As a result, the allocation of cellular resources to specialized genome continuously increases with time. This leads to a deficit of reparative and adaptive capacity in post mitotic cells. Suggestions for future research focus on identifying groups of genes responsible for regulation of growth rate of specialized genome and suppressing ability of the cell division. A better understanding of the relationship between the two parts of the genome will not only help us to manipulate ontogenesis and aging, but will also improve our understanding of cancer development and ontogenesis.

# Introduction

The goal of the present work is to offer the author's view on the causes and mechanisms underlying aging in multicellular organisms. Here, we will consider a generalized process of aging in a multicellular organism. A generic multicellular organism will be taken as the object of the description and analysis in this work.

At present, there is a vast amount of literature dedicated to the subject of aging. We will narrow our review of the current literature on aging by examining several key theories from the standpoint of what has been postulated by their authors and followers.

The early approach to thinking about aging, which we will call the "Wear and Tear" view, postulates that biological systems are inherently unable to oppose the damage due to stochastic chemical or thermodynamic processes (Comfort 1964; Orgel 1963; Szilard1959). The most recognized theory of aging, based on this postulate, is the "Free radical theory". According to this theory, free radicals producing with cell metabolism, constantly damaging cell DNA (Ames et al. 1993; Barja 2002; Beckman and Ames1998; Cristofalo et al. 1998; Fleming et al. 1992; Gallant and Kurland 1997; Gallant and Parker 1997; Gensler and Bernstein1981; Hamilton, et al. 2001; Larsen 1993; Longo and Fabrizio 2002; Lu et al. 2004; Martin et al. 1996; Mori et al. 1998; Parrinello et al. 2003; Sohal et al. 2002; Terman 2001). However, by postulating an inherent imperfection in the construction of living organisms, the authors of this theory ignore, for example, the fact of existence in nature of bacterial species, which have lived for eons while their genomes have remained practically unchanged.

Inherent imperfection of an organism is, similarly, postulated by theories based on mathematical models related to the theory of reliability (Dice et al. 1987; Gavrilov and Gavrilova 2001). These theories explain aging as the product of inevitable errors that occur in the process of the normal operation of any complex system. Still, what such theories overlook is a basic distinguishing feature of the living systems – the ability to repair damage. The theories of aging, which point to the degenerative processes in the various tissues within the organism such as neuroglia (Mattson et al. 2002; Boyko 2007), etc., can also be grouped into the same category. However, here too, it is left unclear, what constitutes the trigger mechanism for such changes in the tissues. In other words, what causes aging? Many other hypotheses can be grouped within the *wear and tear* set of aging theories, in which the disruption of the functionality of various parts of the cell is identified as the cause of aging (Eriksson et al. 2003; Friguet et al. 2000; Harman1972; Marcel Leist et al. 1997; Lutter et al. 2001).

An alternative to the aforementioned subset of theories is a larger group of theories of aging, referred to collectively as evolutionary theories. We shall assign them as theories that postulate the presence in the organism of a genetically predetermined *mechanism of death* or *program of aging*. These theories propose that in the course of evolution, the mechanisms of aging have evolved along with the emergence of the multicellular organisms, out of necessity for their continued evolution. Within this framework, it is worth noting a few works aimed at identifying a "death gene." In these works, the genome is attributed with having a predetermined program of assured destruction of the organism, a sort of a "genetic time bomb," or a programmed disruption of the normal gene function in the cells. Such view of the underlying causes of aging is presented in the classical works of Weissmann 1891; Medawar 1952; and other authors (Friedman and Johnson1988; Gavrilov and Gavrilova 2002; Goldsmith 2004; Guarente and Kenyon 2000; Johnson 1990; Keeley and Bond 1999; Lithgow et al. 1995; Miller 1999; Zwaan et al. 1991).

According to a subset of evolutionary theories known as the theory of antagonistic pleiotropy (Williams 1957), aging occurs because natural selection contributes to the consolidation of the alleles, which have a favorable effect early in life, even if they are unfavorable later on (Campisi 2001; Carlesworth 1994; Carnes and Olshansky 1993; Economos and Lints 1986; Gavrilov and Gavrilova 1999; Itahana et al. 2001; Krtolica et al. 2001; Le Bourg 2001; Partridge et al. 1999; Rose 1991; Westendorp and Kirkwood 1998). Thus, the selection contributing to such allele is realized only in the formative period and is indifferent toward the function of the cell's genes later in life of the individual. This situation, according to the author and the supporters of this view, leads to a cascade of errors and damage accumulation in the genetic machinery and constitutes the cause of aging of the multicellular organisms. Although, until now, a gene or a group of genes that, whether naturally or artificially mutated, would change the life span of the species by an appreciable amount, has not been identified.

Attempts to identify a specific genetic mechanism responsible for aging also include the work devoted to the function of the chromosomal machinery of the cells. This work led to the discovery of the limit on the number of cell divisions, named the Hayflick Limit after its discoverer (Hayflick and Moorhead 1961; Hayflick 1985; 1994; 2000; 2004). In later work, this phenomenon was linked to the chromosomal telomeres (Artandi et al. 2002; Bassham et al. 1998; Benard and Hekimi 2002; Blackburn 2000; Blasco 2005; Collins and Mitchell 2002; Lindsey et al. 1991). In both cases, the authors propose a mechanism that attributes the cause of aging to the limitation on the number of cell divisions. This explanation of aging has two major shortcomings. Firstly, it fails to account for existence of stem cells, which have no such limit on the number of divisions they can undergo at all stages of life. Secondly, this theory applies only to the dividing cells, leaving out the postmitotic cells that form the basis for the functional activity of an organism.

Let us examine in some detail one of the more modern theories of aging, which attempts to unite currently known facts and perspectives: the theory of the Disposable Soma, put forward by Thomas Kirkwood 1977; (Kirkwood and Holliday1979; Kirkwood and Austad 2000). According to Kirkwood, aging is a result of the progressive limitations placed on the energy expenditure of the organism on the repair of the soma, which arise as a result of the competition for resources between reparative and reproductive processes. In Kirkwood's view, the key processes that determine the life span and the rate of aging are controlled by genetic determinants that regulate the supporting functions such as DNA repair genes, antioxidant enzymes, and stress response proteins. Kirkwood argues that, while the process of aging is stochastic in its nature, the aforementioned genes preprogram the life span, as a whole. In his theory, the maximum life span is not defined as a clear time frame but can be changed, for example, by altering the ratio between the rate of damage and the rate of soma maintenance.

In summary, we see that the scientific understanding of the nature and causes of the phenomenon of aging in multicellular organisms consists of two main postulates: *inherent imperfection* of biological systems and the presence of a *program of aging*. The research in these areas is ongoing, though, at present, no one theory provides a satisfactory explanation for what drives the process of aging.

Based on the empirical evidence to date, the main contender for the primary cause of aging is the inability of the cell to repair the damage to its own genome completely. For example, when a result is obtained demonstrating an increase in the DNA damage in the old cells relative to the young cells, a conclusion is made that this is the *cause* of aging. However, this raises a question of whether this really is the cause of aging or is it merely indicative of aging and is a consequence of another process? After all, correlation is not the same as causation. In other words, consequences of unrelated processes may well be mistaken for the cause of the phenomenon in question.

Following such logic, it seems possible to consider infinitely more elementary levels of organization, from the molecular down to quantum mechanical, and still not find the answer to the core question as to what is the cause of aging. It is worth noting, that the theory of inherent imperfection remains attractive as it provides an illusory possibility of fixing the errors of nature thereby obtaining a perfect and, hence, an immortal organism.

The second postulate, namely the presence of a program of aging in multicellular organisms arising from an evolutionary necessity for a mechanism of generational turnover, is similarly plausible. By denying the existence of any weak links in the organism at the cellular level and taking into equal consideration the importance of all its parts, this approach naturally leads to the supposition of a special program directed at aging. Though, let us reiterate, that up until now such a program that would be able to drastically alter the life span of the experimental subject has remained elusive.

## Two Parts of the Cellular Genome

The separation of the cellular genome into two effective functional parts offers a possible means for identifying causes of decline in adaptive and reparative capacity of the individual organism and the consequent aging. These causes can be traced to three key principles:

1. The genome of a multicellular organism is distinct from a unicellular organism in that it consists of two functionally distinct parts. One part, which is more or less common to all cells and organisms, and encodes the basic cellular processes and the other that is specific to each cell type, providing the cells with their specific functionalities.
2. An evolutionary mechanism that ensures the turnover of the population is based on the regulation of the regenerative functions of the organism. Such regenerative functions are the purpose of regulatory mechanisms that ensure a continuing process of evolution or adaptation of the species.
3. There exists a mechanism for the regulation and allocation of resources for the two parts of the genome in the cells of multicellular organisms. This mechanism leads to a gradual increase of expression of the specialized part of the genome and reduces the regenerative capacity of a multicellular organism, leading to the observable age-related changes.

In order to understand these principles better, let us examine the relationship between the organism as a whole and its constituent cells. Let's start with a definition of what an organism is. Is it the sum of the specialized cells participating in a common process or is it a sufficiently independent formation? From current biological knowledge, we know that a multicellular organism is an independent system, with a complex ensemble of many cells, integrated tissues and organs interconnected with the aid of chemical factors. The cells of multicellular organisms are *totipotent* (i.e., they have the genetic potential to be any cell of a given organism) and are equivalent in their genetic information, with the difference defined by the subset of the genes that are expressed. This difference in genome expression, leads to the variety of morphologically and functionally distinct cells in the organism and is based on the phenomenon of *differentiation*.

Interestingly, with the seemingly infinite variety of life forms in the world, only a fairly small number of basic types of cellular structures and fundamental processes exist in multicellular organisms. In principle, there is no difference in the metabolism, biosynthesis, or informational exchange on the cellular level (at least in animals). It can be said that all basic structures of the cells are far more similar than they are different; even multicellular organisms that are far apart on the evolutionary tree have similar organs and systems that function according to universal principles. It is natural to assume, that the similarity in the structural framework results in the convergence of the properties of the biological objects, expressed, as the universality of their basic functions on the cellular level, as well as on the level of the whole organism.

Any process occurring in the organism, including ontogenesis, is primarily determined by the genes of the DNA in each cell. If we consider a multicellular organism from the standpoint of its genome, we observe that its cellular DNA consists of two functionally distinct parts. The first part contains the genes necessary for the cell to define its specialized functions within the multicellular organisms. This part of the genome is present in all cells and provides them with specific functionality, depending on which genes are active in each particular case. We will denote this as *SG* (specialized part of the genome) for brevity. The other part of the genome present in the multicellular organisms is known as the housekeeping genes (Dermitzakis and Clark 2002; Domènec Farré et al. 2007; Lenhard et al. 2003; Tagle et al. 1988; Wray et all 2003; Niewczas E et al. 2000). We will denote it as *HG*. This is the part, which contains the genes that ensure the existence of each individual cell. It can be called a "universal platform" on which evolution and ontogenesis is based in the multicellular organisms. Such division of the genome is justified not only from the standpoint of its function, but also from the point of view of the evolutionary origin of multicellularity (Boldachev 2007).

It is generally agreed that, at the cellular level, the specialized genes of the organism do not participate in the processes of cell life, leaving this role for another part of the cellular genome. The term minimal genome is used in the literature dedicated to the creation of an artificial genome based on the mycoplasmal DNA (Hutchison et al. 2002) and is very close in its definition to housekeeping genes. In fact, the goal of the reproduction of the simplest possible genome of a bacteria itself has led to the creation of this term. By definition, such a genome must contain only the genes essential for existence of an organism.

Adaptational responses connect any biological system with the environment only at the level of the energy-material exchange, while following the laws of thermodynamics. Such processes are well described within the framework of thermodynamic or binary approaches and models. However, we must reiterate here, that the *external environment* is always the object of consumption by the biological system. For any biological object or system, external surroundings are its source of hydrocarbon and protein consumption. At the same time, the external environment produces a constant, nonspecific external influence on its internal parameters. For a biological system, the response to the external stimuli is determined by its internal makeup, represented by its genome. There is no direct relationship to the external environment, producing a closed system. Accordingly, the biological systems do not produce any operational, or binary, effects on the external environment. The internal regulatory processes of the biological systems are completely indifferent to what is happening in the external environment. It is important to note that the regulatory mechanisms on the cellular and organismal levels were not created for intentional interactions with the external environment. They are not included in the range of downstream and upstream interactions, making the events of the external and internal environments causally independent. A detailed description of the properties of such operationally closed systems can be found in studies dedicated to the theory of *autopoiesis* (Maturana and Varela1980).

## Cellular Membranes and the Processes of Aging

At present, it is known that all structural and functional proteins that perform a specialized role are synthesized in the cell, either on the polysomes associated with the membrane, or on the endoplasmic reticulum (ER). At the same time, proteins necessary for the function of the cell itself are synthesized on free polysomes (Ni-Stor and Saez de Cordova 1974). The membrane proteins, which make up the structural and functional basis for the membrane-bound complexes in the cell, are an exception.

Specific structures of the neurons, myocytes (muscle cells), muscle fibers, hepatocyte oxidases, and all secreted proteins, as well as specific proteins in all epithelial tissues, are either associated with, or are parts of, membrane structures. In other words, we can assert that all proteins encoded by the *SG* part of the genome are synthesized on membrane-bound polysomes, whereas *HG* proteins, with the exception of the membrane proteins themselves, are synthesized in the cell by free polysomes.

This situation is biologically justified. It makes sense that the location of synthesis and the location of function are interrelated. It has been experimentally determined that the ratio between the number of free and bound polysomes changes to favor the latter, while in malignant growth, for example, the number of the free polysomes increase. We can conclude that the ratio of free and bound polysomes reflects the ratio of expression of *HG* (free polysomes) and *SG* parts (bound polysomes) of the cellular genome.

Mechanisms exist in the cell to process the proteins that are synthesized within it. In particular, a mechanism, directing proteins to either free or bound polysomes, based on the presence of a lipophilic region in the beginning of the polypeptide chain. Because of this region, the nascent protein with its polysomes attaches to the external cellular membrane or to the endoplasmic reticulum. During the posttranslational processing of the polypeptide, the lipophilic region is cleaved off, and the nascent protein either remains attached to the membrane, is transported through it or, in some cases, is inserted into a membrane-enclosed compartment.

At present, no evidence exists for difference in the rate of transcription of the RNA of *HG*- and *SG*-encoded proteins. Similarly, such evidence does not exist for the difference in the rates of translation. However, there is sufficient experimental evidence suggesting that, at the level of transcription and translation, the rate of operation of the *HG* part of the genome is regulated primarily by a feedback mechanism inherited by the cell from its unicellular ancestors. Activity of the *SG* part, on the other hand, is regulated by mechanism that exists only at the organismal level.

As was previously noted, the ontogenetic program of the organism must pursue two major goals – to reproduce and to produce the generational turnover necessary for the evolution of the species. In this regard, the central part of ontogeny is sexual maturation, a time when an individual fulfills its evolutionarily significant role as a species. Along with cessation of growth of the organism (increase in linear dimensions and mass), a period of sexual maturity begins – maturity of all its functional systems supported by the function of the *SG* part of the genome.

It can be concluded that the rate of development of an individual is determined by the degree of "switching" of the expression levels of the *HG* part of the genome in the cells to the *SG* part, which is necessary for the formation of all functional organs and tissues. This situation accurately reflects the law of ontogenesis, according to which the rate of growth and development is strictly proportional to the rate of degradation. In other words, the slower the growth, the greater the life span. The point of the maximum adaptational stability of an organism occurs during the period of sexual maturity and corresponds approximately to the point of the cessation of growth. Specific, genetically determined, linear dimensions of an organism are attained by a certain number of cell divisions, the limit of which is set by the terminal differentiation of the cell with the subsequent loss of ability to divide, which leads to a gradual increase in expression of *SG* part of the genome.

The early part of ontogenesis is clearly dominated by the expression of the *HG* part of the genome, which provides the necessary growth and development of the individual during this time. In other words, in the course of the first half of the postnatal ontogenesis, a clear advantage is observed of the "production of the means of production" over the "production of the consumables." With growth and development of the organism, its adaptational abilities increase, reaching their maximum by the period of sexual maturity.

We can illustrate the course of these processes with the following diagram 1.

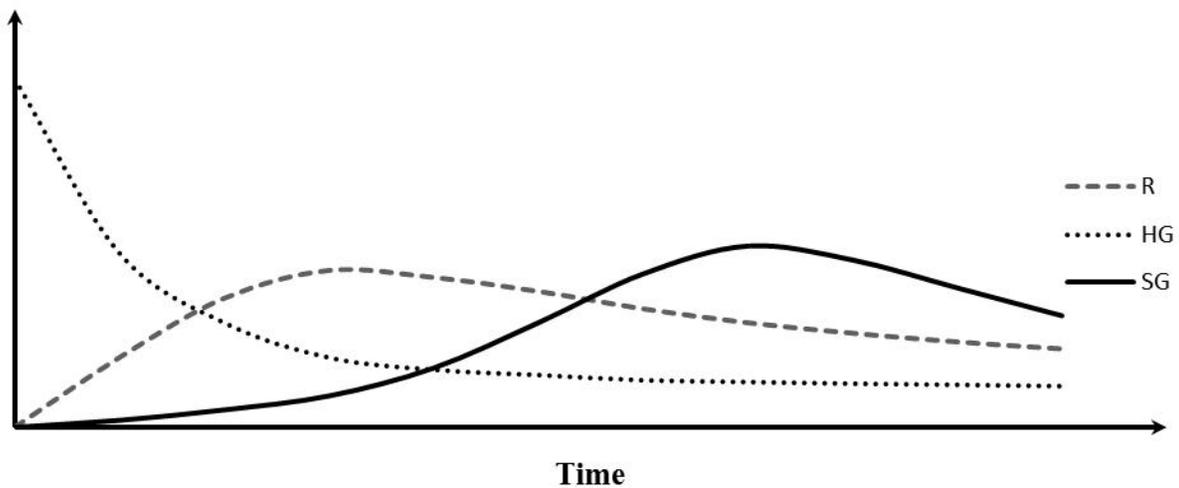

*Diagram 1: Level of HG and SG activity and reparative abilities of an organism over its life span*

The x-axis represents time, and the y-axis is the level of expression of the two parts of the genome – *HG* and *SG*. The dotted curve shows the level of production of *HG* and the solid of *SG*. The dashed line represents the resulting curve—adaptational and reparative abilities of the organism in the course of a lifetime.

As follows from the illustration above, in the course of development, an organism passes the peak of its regenerative abilities in the range of cessation of growth and reaching of sexual maturity. Further increase in the portion of expression of *SG* invariably leads to decline of the synthesis of the products of *HG* and, as a result, the lowering of the adaptational and reparative abilities of an organism.

As we can see, the basic properties of the multicellular organisms, namely their adaptational and reparative abilities, are determined by the ratio of expression of the *HG* part of the cellular genome of the cells themselves and the *SG* part responsible for the functions performed by these cells, which make up the organism with all its functions, as a whole. It is here, at the level of expression of the two parts of the genome that the *competition for resources* arises in the cell.

Recall that for an organism, as a whole, the postmitotic cells, which constitute the basis of a multicellular organism, play the decisive role in these processes. As shown in our previose papers, devoted to mathematical modeling of adaptation processes, they are mainly determined by the properties of regulation. This function in the body is only performed by postmitotic cells. (Salnikov L.V, Bazhenov P.V 2000). These cells determine the level of adaptation abilities and available resources in the organism, which is in turn dependent on which functional part of the genome (*HG* or *SG*) is prioritized. This prioritization is based on the total surface area of ER, which diverts cellular resources and, more importantly, reparative abilities to itself.

What are possible causes of the increase of the ER surface area in the course of aging? One supposition is that following the cessation of cell division, the surface area of intracellular membranes tends to increase by inertia. An objection can be made that the preceding growth of the ER membranes is not necessary to the cell as the increase of their surface area is necessary not *before* but *after* the process of cell division. It is more likely that such process is directly linked to the expression of the *SG* part of genome. We can hypothesize that there is a specific mechanism for the preceding ER growth in the no dividing cells presented in the *SG* part of the genome. Such hypothesis is logical, as such mechanism is necessary for the *SG* part to perform its function. However, it is perhaps more likely that the *SG* part simply has the ability to stimulate, or *control*, the surface area of ER, as needed by employing the *HG* part of the cellular genome for this purpose.

However, if the regulation of the rate of ER synthesis occurs in the specialized cells according to the principles of feedback mechanisms, why does the level of this synthesis not stop at the equilibrium point instead of continuing to grow with the consequences we have already discussed. That is the essential question for the analysis of causes of aging mechanisms. Recall that one of the basic properties of any open thermodynamic system is constant fluctuation, a delicate balance of their parameters. As was noted earlier, due to the huge number of connections in such system, we can talk about a continuous adaptation to adaptational responses occurring in the cell itself. Such a state, even without significant external stresses, continuously stimulates all functions of the organism as a system, creating a situation of constant demand for the strengthening of the adaptational response. At the same time, as was previously shown, the internal needs of the cell are controlled only by its own internal connections.

Processes at the organismal level recognize their cellular basis as simply a part of the external environment without including it in their regulatory system. The result is that all processes of the replication of the synthetic machinery of the cell and all reparative processes are inconsequential or invisible for the multicellular organism as whole, as a higher-level system. Importantly, this situation becomes fatal only in the case when it arises within a space limited by the physical dimensions of the cell. This is exactly the situation that occurs upon the completion of the growth of the organism and the increase of the size of the postmitotic cells.

Again, after reaching sexual maturity, a gradual increase of the load on the *SG* part of the cellular genome is observed, leading to a steady decline of the capacity of the *HG* part. In this way, the reciprocal, competitive relationship between the expression of *HG* and *SG* parts of the genome of the differentiated cells gradually shifts toward the expression of the *SG* part. This situation is also associated with the limitation of the physical growth of the cells themselves. In this case, the reduction of the adaptational and reparative potentials of the cell occurs as a result of gradually increasing portion of the expended hydrocarbon and protein resources, being diverted to synthesis of membrane complexes and the functional products associated with the products of the *SG* part. The result is a concurrent decrease of the volume of the available cytoplasm and free polysomes, where the synthesis of the structures encoded by the *HG* part occurs. This process is self-accelerating in nature, as the number of attachment spaces for bound polysomes, synthesizing proteins encoded by the *SG* part of the genome, increases as a function of the increase in the surface area of ER.

This analysis explains a series of theories of aging, linking the phenomenon with age-related damage of the genetic apparatus of the cells of an organism. Concluding this section, we return to the phenomenon of ontogenesis itself. Based on the foregoing argument with regard to age-related processes, we can draw the following conclusion: ontogenesis determines *only* the rate of formation and growth of an individual, leaving the natural degradation and death of an individual to the method of organization, or the principles of construction of any multicellular organism.

# Conclusions and Perspectives

The approach to the multicellular organism, as a complex of two functionally independent genetic systems, leads to several predictions. In the course of ontogenezis, runs a special mechanism for "capturing" the control of cell division. In other words, the cells lose their ability to control the process of self-division. *SG* part of the genome not only becomes the sole regulator of cell division, but also controls the operation of postmitotic cells.

The major findings presented here result from the proposed view of the mechanisms of aging. First, we note that the particular "universal" mechanism of aging has not yet been identified. Instead, there are the two distinct parts of the organism cell genome, *HG* and *SG*. The fact that these two parts were designed to function independently leads to competition for cellular resources. This competition leads to all manifestations of the phenomenon of aging.

The perspective proposed here, rationalizes offers a new approach to studying and understanding the process of aging. It becomes clear that the ratio between the duration of the growth phase and the life span of an organism is based on the fact that following the end of growth (ontogenetic program execution), the output of the *SG* genome begins to prevail in all postmitotic cells at a rate proportional to the rate of growth of the organism. As a result, physically limited resources of the cells begin to gradually shift from reparative capacity to execution of external functions. This explains the increase in the amount of cellular damage by the free radicals.

Based on this approach, the main thrust of the research directed at uncovering aging mechanisms, should be the question: how is the ontogenetic program implemented, starting with the earliest stages of embryogenesis. During this time, the genes that "intercept" the regulation of the cell division process, start functioning. Identification of specific genes that control the cell division cycle in the body will help to understand how the processes of tissue regeneration work. At the same time, the analysis of these genes will help understand the mechanisms of cancer. Ontogenetic regulation of the cell division cycle program, has been shown to be directly connected with the development of the aging process. Identification and study of genes that participate in the mechanisms, that regulate the rate of growth and development in multicellular organisms, will enable not only to affect the rate of growth of the organism, but also to control the processes of aging.